\input amstex
\documentstyle{amsppt}
\document
\magnification=1200
\NoBlackBoxes
\nologo
\pageheight{18cm}

\bigskip

\centerline{\bf HOLOGRAPHY PRINCIPLE} 

\smallskip

\centerline{\bf AND ARITHMETIC OF ALGEBRAIC CURVES}

\medskip

\centerline{\bf Yuri I. Manin, Matilde Marcolli}

\medskip

\centerline{\it Max--Planck--Institut f\"ur Mathematik, Bonn, Germany}

\bigskip

{\bf Abstract.} According to the holography principle
(due to G.~`t Hooft, L.~Susskind, J.~Maldacena, et al.), 
quantum gravity and string theory
on certain manifolds
with boundary can be studied in terms of a conformal field theory on
the boundary. Only a few 
mathematically exact results corroborating this exciting program
are known. In this paper we interpret from this perspective
several constructions which arose initially
in the arithmetic geometry of algebraic curves. We show that
the relation between hyperbolic geometry and Arakelov geometry
at arithmetic infinity involves exactly the same geometric data 
as the Euclidean AdS${}_3$ holography of black holes. Moreover, in the
case of Euclidean AdS${}_2$ holography, we present some results on
bulk/boundary correspondence where the boundary is a non--commutative
space.

\bigskip

\centerline{\bf \S 0. Introduction}

\medskip

{\bf 0.1. Holography principle.} Consider a manifold $M^{d+1}$
(``bulk space'') with boundary $N^d$. The holography principle
postulates the existence of strong ties between certain field
theories on $M$ and $N$ respectively. For example, in the actively
discussed Maldacena's conjecture ([Mal], [Wi]), $M^{d+1}$ is  
the anti de Sitter space AdS${}_{d+1}$ (or 
AdS${}_{d+1}\times S^{d+1}$), $N^d$ its conformal boundary.
On the boundary one considers the large $N$ limit of  a conformally
invariant theory in $d$ dimensions, and on the bulk space supergravity
and string theory (cf.~e.g. [AhGuMOO], [Mal], [Suss], ['tH], [Wi],
[WiY]). 

\smallskip

The holography principle was originally suggested by `t Hooft
in order to reconcile unitarity with gravitational collapse.
In this case $M$ is a black hole and $N$ is the event horizon.
Thus the bulk space should be imagined as (a part of) space--time.

\smallskip

There are other models where the boundary can play the role of
space--time (Plato's cave picture), with 
the bulk space involving an extra dimension (e.~g. the
renormalization group scale) and a Kaluza--Klein type reduction
[AlGo], and ``brane world scenarios'' where one models
our universe as a brane in higher dimensional space--time, with
gravity confined to the brane. 

\smallskip

In this paper we consider first of all a class of Euclidean AdS${}_3$
bulk spaces which are quotients of the real hyperbolic 3--space
$\bold{H}^3$ by a Schottky group. The boundary (at infinity) of such a
space is a compact oriented surface with conformal structure,
which is the same as a compact complex algebraic curve. Such spaces are
analytic continuations of known (generally rotating) Lorentzian
signature black hole solutions,  and they were recently studied from
this perspective by K.~Krasnov (cf. [Kr1]--[Kr4].)

\medskip

{\bf 0.2. Arithmetic geometry at infinity.} Consider a
projective algebraic curve $X$ defined, say, over the field
of rational numbers $\bold{Q}$. It can be given by equations
with integer coefficients which defines a scheme $X_{\bold{Z}}$,
``arithmetical surface''. $X$ itself is the generic fiber of
the projection $X_{\bold{Z}}\to \roman{Spec}\,\bold{Z}.$
Finite points of the ``arithmetic curve'' $\roman{Spec}\,\bold{Z}$ are
primes $p$, and 
the closed fibers of $X_{\bold{Z}}$ at finite distance are the
reductions $X_{\bold{Z}}\,\roman{mod}\, p$. One can also consider 
infinitesimal neighborhoods of $p$ and the respective fibers
which are simply reductions of $X_{\bold{Z}}$ modulo powers $p^n$.
The limit of such reductions as $n\to\infty$ 
can be thought of as a $p$--adic completion of  $X_{\bold{Z}}.$

\smallskip

A geometric analog of this picture is an algebraic surface fibered over
an affine line (replacing $\roman{Spec}\,\bold{Z}$.) We can complete
the affine line to the projective one by adding
a point at infinity, and extend the fibered surface by adding
a closed fiber at infinity. If we want to imitate this
in the arithmetic case, we should add somehow ``the arithmetic
infinity'' to $\roman{Spec}\,\bold{Z}$ and enhance 
the geometry of $X$ by appropriate structures. 

\smallskip

It was long known that the arithmetic infinity itself is represented 
by the embedding $\bold{Q}\to \bold{C}$ and considering the complex 
absolute value
on an equal footing with $p$--adic valuations.
In his paper [Ar] S.~Arakelov demonstrated that Hermitian geometry of
$X_{\bold{C}}$ constitutes an  analog of
$p$--adic completions of $X_{\bold{Z}}$. In particular, Green's functions
for appropriate metrics provide intersection indices
of arithmetic curves at the infinite fiber.
Arakelov's arithmetic geometry
was since then tremendously developed and generalized
to arbitrary dimensions. 

\smallskip

One aspect of $p$--adic geometry was, however, missing
in Arakelov's theory of arithmetical infinity: namely, 
an analog of the closed fiber
$X_{\bold{Z}}\,\roman{mod}\, p$ and the related picture
of reductions modulo powers of $p$ approximating the $p$--adic limit.

\smallskip

In Manin's paper [Man2] it was suggested that this missing structure
can be modeled by choosing a Schottky uniformization
of $X(\bold{C})$ and treating this Riemann surface
as the conformal boundary of the respective
handlebody obtained by factoring out $\bold{H}^3$ with respect to the
Schottky group. Comparing this structure with 
the $p$--adic case, one should keep in mind that
only curves with maximally degenerate reduction (all components of genus zero)
admit a $p$--adic Schottky uniformization (Mumford's theory).
Thus we imagine ``the reduction modulo arithmetic infinity''
to be maximally degenerate: a viewpoint which is supported by other
evidence as well.

\smallskip

We see thus that the $\infty$--adic geometry at arithmetic infinity,
developed in [Man2], involves exactly the same geometric data 
{\it bulk space/boundary}
as the Euclidean AdS${}_3$ holography of black holes.
Moreover, Arakelov's intersection indices are
built from Green's functions, which form the basic building blocks
for Polyakov measures as well as the correlation functions of bosonic
and fermionic field theories on $X$ (see [ABMNV], [Man1], [Fay],
[FeSo].) 

\smallskip

In the first section of this  paper we demonstrate that the
expressions for these 
Green functions in terms of the geodesic configurations in the handlebody 
given in [Man2] can be nicely interpreted in the spirit
of the holography principle. 

\smallskip

A recent attempt to generalize [Man2] to higher dimensions
is due to A.~Werner ([We]). It would be interesting
to discuss her construction as a case of holography.

\medskip  

{\bf 0.3. Modular curves and non--commutative boundary.} 
The second section is dedicated to the holography
in 1+1 dimensions which we recognize in the approach to
the theory of modular curves developed, in particular,
in [ManMar]. In this case $\bold{H}^3$ is replaced
by the upper complex half--plane $\bold{H}^2$, and a Schottky
group by a subgroup $G$ of the modular group. The most
interesting new feature is that the boundary of the quotient
space considered in [ManMar] is a non--commutative space:
it is the quotient $G\backslash\bold{P}^1(\bold{R})$ treated
as a crossed product in the style of Connes. This might be of
interest, because non--commutative boundaries of moduli
spaces (e.~g. that of instantons) play an increasingly
important role in physics considerations.

\smallskip 

In particular, we argue that one reason why little is known 
on $AdS_{1+1}$ holography, unlike the much better understood 
case of $AdS_{2+1}$, is that a treatment of holography for 
$AdS_{1+1}$ and its Euclidean counterpart $\bold{H}^2$ 
should take into account the presence of non--commutative 
geometry at the boundary. 

\medskip

{\it 0.4. Acknowledgment.} We are grateful to Alain Connes who suggested
the authors to look at [Man2] from the perspective
of the holography principle. We also thank Kirill Krasnov for several
useful and encouraging comments.

\bigskip

\centerline{\bf \S 1. Handlebodies as holograms}

\medskip

In this section we review the basic notions of the
boundary and bulk geometry and function theory
in the context of Schottky uniformization.
Then we state and interpret the main formulas of [Man2]
in the light of the holography principle.

\smallskip

{\bf 1.1. Green's functions on Riemann surfaces.} Consider
a compact non--singular complex Riemann surface $X$ and a divisor
$A=\sum_x m_x(x)$ on it with support $|A|$. If we choose a positive
real--analytic 
2--form $d\mu$ on $X$, we can define the Green function
$g_{\mu ,A}=g_A$ as a real analytic function on
on $X\setminus |A|$. It is uniquely determined by the following
conditions.

\smallskip

(i) {\it Laplace equation}:
$$
\partial\bar{\partial}\,g_A=\pi i\,(\roman{deg} (A)\,d\mu -\delta_A)
$$
where $\delta_A$ is the standard $\delta$--current
$\varphi\mapsto \sum_x m_x\varphi (x).$

\smallskip

(ii) {\it Singularities:} if $z$ is a local parameter in a neighborhood of
$x$, then $g_A - m_x\roman{log}\,|z|$ is locally
real analytic.

\smallskip

(iii) {\it Normalization:} $\int_X g_Ad\mu =0$.

\smallskip

Let now $B=\sum_y n_y(y)$ be another divisor, $|A|\cap |B|=\emptyset .$
Put $g_{\mu}(A,B):=\sum_y n_yg_{\mu , A}(y).$ This is
a number, symmetric and biadditive in $A,B$.

\smallskip

Generally, $g_{\mu}$ depends on $\mu$. However, if $\roman{deg}\,A=
\roman{deg}\,B=0,$ $g_{\mu,A,B}$ depends only on $A,B$.
Notice that, as a particular case
of the general K\"ahler formalism, to choose
$d\mu$ is the same as to choose a real analytic Riemannian metric on $X$
compatible with the complex structure. This means that $g_{\mu}(A,B)=
g(A,B)$
are conformal invariants when both divisors are of degree zero. 
If moreover $A$ is the divisor of a meromorphic function $w_A$,
then
$$
g(A,B)=\roman{log}\,\prod_{y\in |B|}|w_A(y)|^{n_y} =\roman{Re}\,
\int_{\gamma_B}\frac{dw_A}{w_A} 
\eqno(1.1) 
$$
where $\gamma_B$ is a 1--chain with boundary $B$. This is
directly applicable to divisors of degree zero on the Riemann sphere
$\bold{P}^1(\bold{C}).$ 

\smallskip
This formula admits also a generalization to arbitrary $A,B$
of degree zero on a Riemann surface of arbitrary genus.
The logarithmic differential $dw_A/w_A$ must be replaced by
the differential of the third kind $\omega_A$ with pure
imaginary periods and residues $m_x$ at $x$. Then
$$
g(A,B)=\roman{Re}\,
\int_{\gamma_B}\omega_A \, . 
\eqno(1.2) 
$$
If we drop the degree zero restriction, we can
write an explicit formula for the basic Green's function
$g_{\mu,x}(y)$ via theta functions in the case when $\mu$
is {\it the Arakelov metric} constructed with the help
of an orthonormal basis of the differentials of the first kind.
For a characterization of Arakelov's metric
in a physical context, see [ABMNV], pp. 520--521.

\medskip

{\bf 1.1.1. Field theories on a Riemann surface $X$.} Green's functions
appear in explicit formulas for correlators
of various field theories, insertion formulas, and Polyakov
string measure. In [ABMNV] they are used in order to
establish the coincidence of certain correlators
calculated for fermionic, resp. bosonic fields on $X$
(bosonization phenomenon.) See [Fay] for a thorough mathematical
treatment.

\medskip

{\bf 1.2. Green's functions and bulk geometry: genus zero case.}
In this subsection $X$ is the Riemann sphere $\bold{P}^1(\bold{C})$.
It is convenient to start with a coordinate--free description of all
basic objects.  

\smallskip

Choose a two--dimensional complex vector space $V$ and define
$X=X_V$ as the space of one--dimensional vector subspaces
in $V$. Define the respective bulk space as a three--dimensional
real manifold $\bold{H}^3=\bold{H}_V$ whose points are
classes $[h]$ of hermitian metrics $h$ on $V$ modulo dilations:
$h\cong h'$ iff $h=\rho h'$ for some $\rho >0$. Clearly, $PGL(V)$
acts on $\bold{H}_V$ and $X_V.$ The stabilizer of any $[h]$
is isomorphic to $SU(2)$. Any point $[h]$ defines a unique
K\"ahler metric on $X_V$ which is stabilized by the same
subgroup as $[h]$ and in which the diameter of $X_V$ equals one.
This metric, in turn, determines a volume form $d\mu=d\mu_{[h]}$
on $X_V.$

\smallskip

The bulk space $\bold{H}_V$ has a natural metric: the distance between
$[h]$ and $[h']$ is the logarithm of the quotient of volumes
of unit balls for $h$ and $h'$, if one ball is contained in the other and their
boundaries touch.  In fact, $\bold{H}_V$ becomes the
hyperbolic three--space of constant curvature $-1$.
Its conformal infinity $X_V$ can be invariantly described as the space
of (classes of) unbounded ends of oriented geodesics.

\smallskip

We will now give a bulk space interpretation of
two basic Green's functions $g((a)-(b),(c)-(d))$
and $g_{\mu}(z,w)$, where $d\mu$ corresponds to a point
$u\in\bold{H}_V$ as explained above. To this end, introduce the following
notation from [Man2]. If $a,b\in \bold{H}_V\cup X_V,$ 
$\{a,b\}$ denotes the geodesic joining $a$ to $b$ and oriented in this
direction. For a geodesic $\gamma$ and a point $a$, $a*\gamma$ is the point
on $\gamma$ at which $\gamma$ is intersected by the geodesic $\delta$
 passing through $a$
and orthogonal to $\gamma$. In particular, the distance from
$a$ to $\gamma$ is the distance from $a$ to $a*\gamma$.
If two points $p,q$ lie on an {\it oriented} geodesic $\gamma$,
we denote by $\roman{ordist}\,(p,q)$, or else $\ell_{\gamma}(p,q)$,
the respective
oriented distance.

\smallskip

\proclaim{\quad 1.2.1. Lemma} We have
$$ 
g((a)-(b),(c)-(d)) = -\roman{ordist}\, \left( a*\{c,d\}, b*\{c,d\}
\right), 
\eqno(1.3) 
$$
$$ 
g_\mu (p,q) = \log \frac{e^{1/2}}{\cosh \roman{dist}\left( u,\{ p,q
\} \right) }. 
\eqno(1.4) 
$$
\endproclaim

\smallskip

The following Fig.1 illustrates the configurations of geodesics
involved.

\bigskip

\input epsf
\midinsert
$$\centerline{\hbox{\epsffile{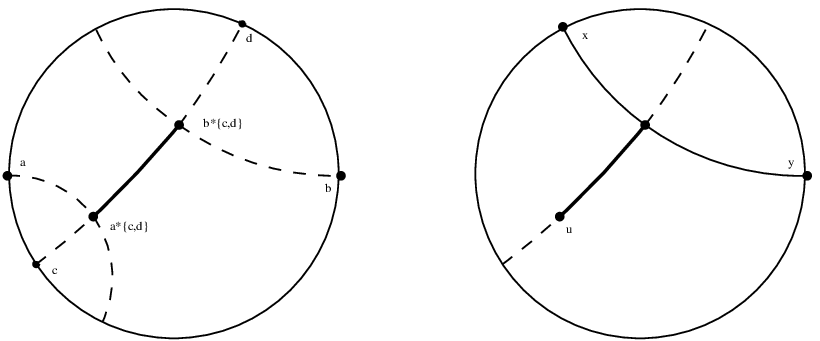}}}$$
\endinsert

\centerline{Fig. 1}

\pagebreak

We invite the reader to compare these configurations
to the Feynman diagrams in [Wi] illustrating
propagation between boundary and/or interior points.

\smallskip

To check, say, (1.3), it is convenient to introduce the standard coordinates
$(z, y)$ identifying $\bold{H}$ with $\bold{C}\times\bold{R}_+$.
Both sides of (1.3) are $PGL(V)$--invariant. Hence it suffices to
consider the case when $(a,b,c,d)=(z,1,0,\infty )$ in $\bold{P}^1(\bold{C}).$
Then $\{c,d\}=\{0,\infty\}$ is the vertical coordinate semi--axis,
and generally in $(z,y)$ coordinates of $\bold{H}^3$ we have
$$
a*\{c,d\}=(0,|z|),\quad b*\{c,d\}=(0,1),
$$
$$
\roman{ordist}\,((0,|z|), (0,1))=-\roman{log}\,|z|.
$$
On the other hand, using the notation of (1.1), we obtain
$$
g((a)-(b),(c)-(d))=\roman{log}\,\frac{|w_{(a)-(b)}(c)|}{|w_{(a)-(b)}(d)|}=
\roman{log}\,|z|.
$$
The middle term of this formula involves the classical cross--ratio of four points on a projective line, for which it is convenient
to have a special notation:
$$
\langle a,b,c,d\rangle := \frac{w_{(a)-(b)}(c)}{w_{(a)-(b)}(d)} \,.
\eqno(1.5)
$$
\smallskip

It is interesting to notice that not only the absolute value, but
the argument of the cross--ratio (1.5) as well admits
a bulk space interpretation:
$$
\roman{arg}\,\frac{w_{(a)-(b)}(c)}{w_{(a)-(b)}(d)} =
-\psi_{\{c,d\}}\,\left( a, b \right), 
\eqno(1.6) 
$$
Here we denote by $\psi_{\gamma}(a,b)$ the oriented angle
between the geodesics joining $a*\gamma$ to $a$ and $b*\gamma$ to $b$,
which can be measured after the parallel translation to, say, $a$.
For a proof of this and other details we refer to [Man2], Prop. 2.2.

\smallskip

This expression is relevant in at least two contexts. First, it
shows how the characteristics of rotating black holes are
encoded in the complex geometry of the boundary
(cf. (1.8) below for the genus 1 case). Second,
it demonstrates that our formulas for the Green functions 
$g(A,B)$ given below
can be refined to provide the bulk space avatars of
the complex analytic expressions such that $\roman{exp}\,g(A,B)$
is the modulus squared of such expression. This is
the well known phenomenon of holomorphic factorization.

\smallskip

We will now introduce a Schottky group $\Gamma$ acting upon
$\bold{H}^3 \cup \bold{P}^1(\bold{C})$
and consider the respective quotient spaces. The boundary will become
a complex Riemann surface $X(\bold{C})$, whose genus equals to number
of generators of $\Gamma$, and the bulk space turns into
a handlebody $\Gamma\backslash \bold{H}^3$ ``filling'' this surface. 
\smallskip

The boundary/bulk expressions
for degree zero Green's functions and related quantities
will be obtained from (1.3) with the help of an appropriate averaging
over $\Gamma$. The geodesic configuration involved
in the right hand side of (1.3) will have to be supplemented
by its $\Gamma$--shifts and then projected into the handlebody.
After such a projection, however, an expression like
$a*\gamma$ will have to be replaced by an infinite sum over all
geodesics starting, say, at a boundary point $a$ and crossing
$\gamma$ orthogonally. Interpreting distances between such points
involved in (1.3) also becomes a trickier business:
the geodesic along which we measure this distance has
to be made explicit. We will provide the details
for the genus one case in 1.3 below. After gaining some experience,
we can restrict ourselves to working in
the covering bulk space $\bold{H}^3$: it is well known that
the geometry of non--simply connected spaces is best described
in terms of the universal cover and its group of deck transformations.
In 1.4 we explain this geometry for genus $\ge 2$ case.

\bigskip

{\bf 1.3. Genus 1 case and Euclidean BTZ black holes.}
Ba\~{n}ados--Teitelboim--Zanelli  
black holes ([BTZ]) are asymptotically AdS space--times which are
obtained by global identifications of $AdS_{2+1}$ by a discrete group 
of isometries $\Gamma$ generated by a single loxodromic element. 

\smallskip

The group of isometries of $AdS_{2+1}$ is $SO(2,2)$
as can be seen by considering the hyperboloid model of anti de Sitter
space $-t^2 -u^2 +x^2 +y^2 = -1$ in $\bold{R}^{2,2}$. 

\smallskip

The non--rotating case (see [ABBHP], [Kr1]) corresponds to
the case where the group $\Gamma$ lies in a diagonal $SO(2,1) \cong
PSL(2,\bold{R})$ in $SO(2,2)$. In this case, there is a surface of time
symmetry. This $t=0$ slice is a two--dimensional Euclidean signature 
space with constant negative curvature, hence it has the geometry of
the real hyperbolic plane $\bold{H}^2$. The fundamental
domain for the action of $\Gamma$ on the $t=0$ slice is given by a
region in $\bold{H}^2$ bounded by two non--intersecting
infinite geodesics, and the group $\Gamma$ is generated by the
element of $PSL(2,\bold{R})$ that identifies the two non--intersecting
geodesics in the boundary of the fundamental domain, creating a
surface with the topology of $S^1 \times \bold{R}$.
The BTZ black hole is then obtained by evolving
this $t=0$ surface in the time direction in $AdS_{2+1}$, until it
develops singularities at past and future infinity. 
The time evolution of the two geodesics in the boundary of the
fundamental domain gives geodesic surfaces that are joined at the past
and future singularities. The geodesic arc realizing the path of
minimal length between the two non--intersecting geodesics is
the event horizon of the BTZ black hole (see [ABBHP], [BTZ], [Kr1]
for further details).

\smallskip

The Euclidean analog of the BTZ black hole is
given by realizing the $\bold{H}^2$ slice as a hyperplane in
$\bold{H}^3$ and ``evolving'' it by continuing the geodesics in
$\bold{H}^2$ to geodesic surfaces in $\bold{H}^3$. This produces a 
fundamental domain of the form illustrated in the Fig 2. 

\bigskip

\bigskip

\bigskip

\input epsf
\midinsert
$$\centerline{\hbox{\epsffile{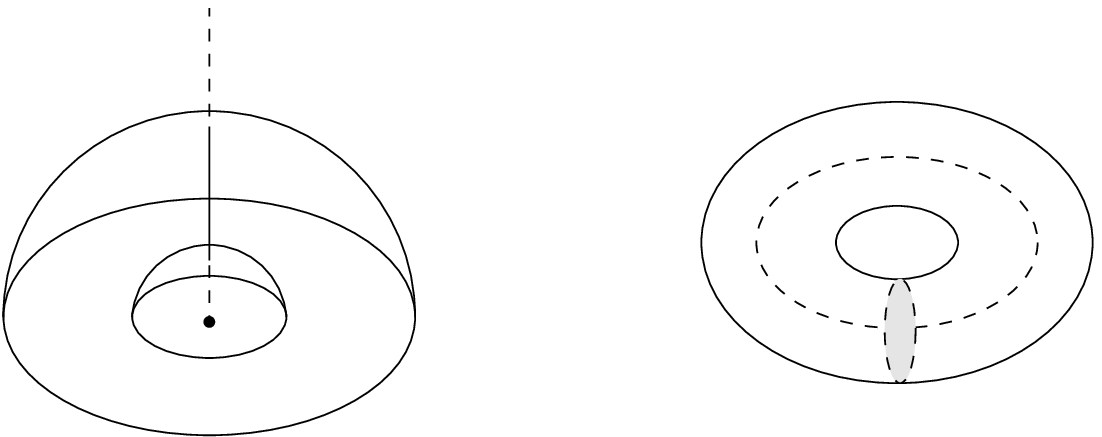}}}$$
\endinsert

\centerline{Fig. 2}

\bigskip

The group $\Gamma\cong q^{\bold{Z}}$ is a Schottky group of rank one
in $PSL(2,\bold{C})$, generated by the choice of an 
element $q\in\bold{C}^*$, $|q|<1$. It acts on $\bold{H}^3$ by 
$$ 
\left(\matrix q^{1/2} & 0
\\ 0 & q^{-1/2} \endmatrix\right) (z,y) = (qz, |q|y). 
\eqno(1.7) 
$$
The quotient ${\Cal X}_q = \bold{H}^3 / (q^{\bold{Z}})$ is a solid
torus with a hyperbolic structure and with the Jacobi uniformized
elliptic curve $X_q(\bold{C}) = \bold{C}^*/(q^{\bold{Z}})$ as its
boundary at infinity. The fundamental domain depicted on Fig. 2
is $|q|^2<|z|\le 1$, $|q|^2<|z|^2+y^2\le 1.$

\smallskip

The physical meaning of $q$ is clarified by the following expression:
$$ 
q= \exp\left( \frac{2\pi ( i|r_-| - r_+ ) }{\ell} \right),
\eqno(1.8) 
$$ 
where the parameters $r_{\pm}$ depend on mass $M$ and angular momentum
$J$ of the black hole,
$$ 
r_{\pm}^2 = \frac{1}{2} \left( M\ell \pm \sqrt{ M^2\ell^2+
J } \right), 
$$ 
and $\ell$ determines the cosmological constant
$\Lambda =-1/\ell^2$  and normalizes the metric as
$$
 ds^2 = \frac{\ell^2}{y^2} (|dz|^2 + dy^2).
$$
This can be seen by writing 
the coordinates in the upper half space model of
$\bold{H}^3$ in terms of Schwarzschild coordinates $(r,\tau,\phi)$
with Euclidean time $\tau$, 
$$ 
z = \left(\frac{r^2-r_+^2}{r^2-r_-^2}\right)^{1/2} \exp\left(
\left( \frac{r_+}{\ell} \phi - \frac{|r_-|}{\ell^2}\tau \right) +
i \left( \frac{r_+}{\ell^2} \tau + \frac{|r_-|}{\ell} \phi \right)
\right), 
$$   
$$ y = \left(\frac{ r_+^2 - r_-^2}{r^2 - r_-^2}\right)^{1/2}
\exp\left( \frac{r_+}{\ell} \phi - \frac{|r_-|}{\ell^2}\tau \right). 
$$
The transformation (1.6) can then be written as
$$ 
(z,y) \mapsto ( e^{2\pi ( i|r_-| - r_+ )/\ell}\, z , \, e^{-2\pi r_+
/\ell} \, y ). 
$$
This was already observed in [BKSW]. For $r_- \neq 0$,
that is, not purely real $q$,
the quotient space $X_q(\bold{C})$ represents a spinning black hole.
We normalized our coordinates so that $\ell =1.$ 

\medskip

{\bf 1.3.1. Determinant of the Dirac operator and Green's function.}
There are explicit formulas  
in terms of theta functions for the
determinant of the Dirac operator 
twisted with a flat bundle on an elliptic curve. For a parameterized
family of Dirac operators 
$D_P$, with $P$ a Poincar\'e line bundle, whose restriction to a
fiber $X_q$ over $L\in Pic^0(X_q)$ is isomorphic
to $L$, it is proved in [RS] that, up to a constant phase, we have 
$$
\det D_P(q;u,v) = q^{\frac{B_2(v)}{2}} \prod_{n=1}^\infty \left(1-q^{n-v}
e^{2\pi i u}\right)\left(1-q^{n+v-1} e^{-2\pi i u}\right), 
\eqno(1.9)
$$
with $B_2(v)=v^2-v+1/6$ the second Bernoulli polynomial.
It is shown in [AMV] that (1.9) is the operator product expansion of
the path integral for fermions on the elliptic curve $X_q$.

On the other hand, the Arakelov Green function on 
$X_q$ is essentially the logarithm of the
absolute value of this expression:
$$ 
g(z,1)= \log \left( |q|^{B_2(\log |z|/\log |q|)/2} |1-z| \,
\prod_{n=1}^\infty | 1- q^n z | \, \, |1-q^n z^{-1} |
\right)
\eqno(1.10) 
$$ 
(see [Man2], (4.6)). 

To interpret various terms of (1.10) via geodesic configurations, we
use (1.3) and (1.5) for various choices of the cross--ratio,
for example, $|x|=|\langle x,1,0,\infty\rangle|$,
$|1-x|=|\langle x,0,1,\infty\rangle|$. More precisely,
we introduce the following notation:

\smallskip

$\bullet$ $\{0,\infty\}$ in $\bold{H}^3$ becomes the closed
geodesic $\gamma_0$ in the solid torus $\Cal{X}_q$. Its length
is $l(\gamma_0)=-\roman{log}\,|q|$ (cf. (1.6).)

\smallskip

$\bullet$ Choose a point $x$ on the elliptic curve $X_q$ and denote
by the same letter $x$ its unique lift to $\bold{C}$
satisfying $|q|<|x|\le 1.$ In particular, 1 denotes both the number
and the identity point of $X_q$. 

\smallskip

$\bullet$ Denote by $\bar{x}$ the point $x*\{0,\infty\}$
and also its image in $\gamma_0$. Similarly, denote by
$\bar{1}=1*\{0,\infty\}=(0,1)\in\bold{H}^3$ and the respective point in 
$\gamma_0$.

\smallskip

Fig. 3a depicts the relevant configurations: 

\pagebreak

\input epsf
\midinsert
$$\centerline{\hbox{\epsffile{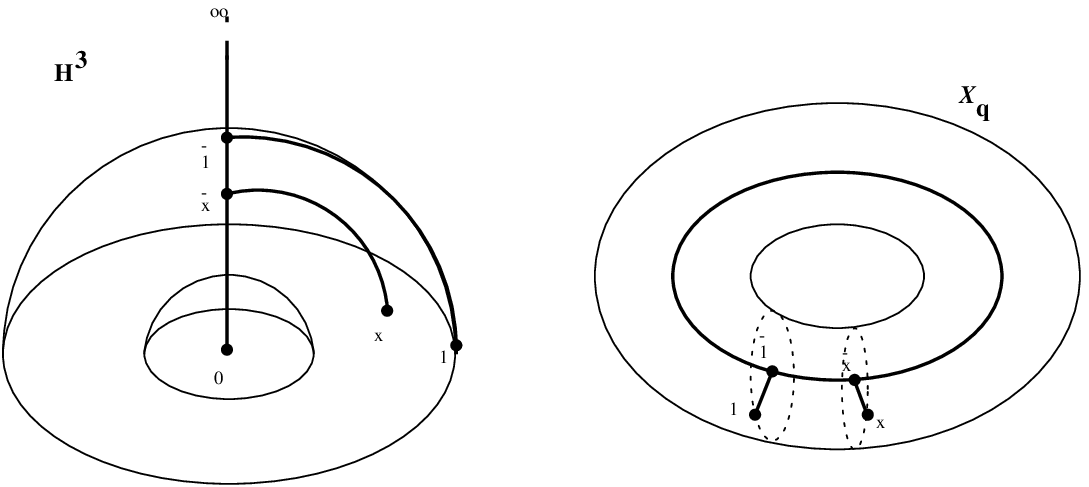}}}$$
\endinsert

\centerline{Fig. 3a}

\smallskip

$\bullet$ Denote the image of $\{1,\infty\}$ by $\gamma_1$.
This is the geodesic starting at the boundary identity
point and having $\gamma_0$ as its limit cycle
at the other end. (As was explained in [Man2],
this is one of the avatars of ``reducing 1 modulo powers
of arithmetic infinity''.) Denote by $\bar{0}$
the point $0*\{1,\infty\}$, and also its image
in the solid torus.

\smallskip

$\bullet$ Finally, put $\bar{x}_n=q^nx*\{1,\infty\}$,
and denote its image in $\gamma_1$ by the same letter.
Similarly, $\tilde{x}_n=q^nx^{-1}*\{1,\infty\}$ (cf. Fig. 3b.)

\bigskip

\input epsf
\midinsert
$$\centerline{\hbox{\epsffile{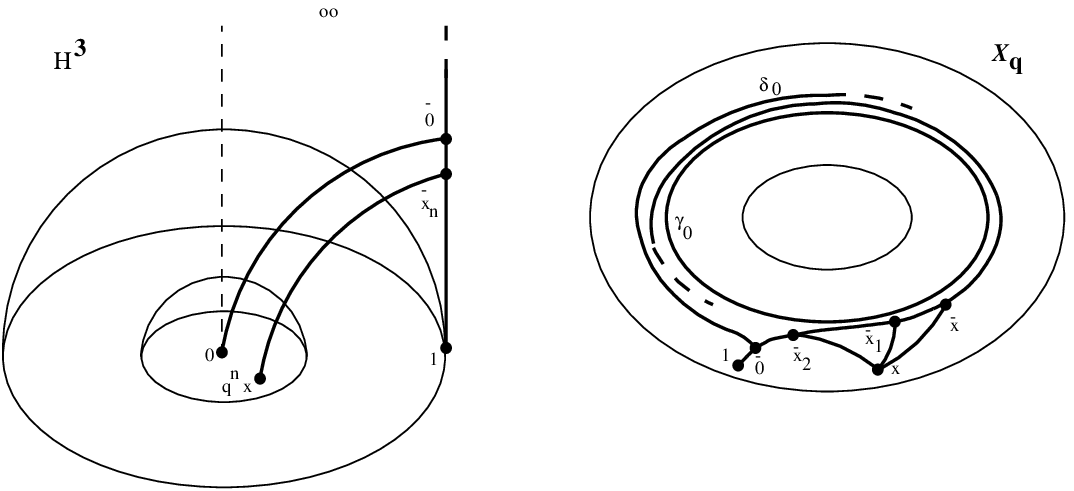}}}$$
\endinsert

\centerline{Fig. 3b}
\pagebreak

With this notation, we have:

\bigskip

\proclaim{\quad 1.3.2. Proposition}  Let $g(u,v)$ be the basic 
Green function with respect to the invariant measure
of volume 1. Then $g(u,v)=g(uv^{-1},1)$, and
$$
g(x,1)=-\frac{1}{2} l(\gamma_0)\, B_2\left(
\frac{\ell_{\gamma_0}(\bar{x},\bar{1})}{l(\gamma_0)}\right)
+\sum_{n\ge 0} \ell_{\gamma_1}(\bar{0},\bar{x}_n)
+\sum_{n\ge 1} \ell_{\gamma_1}(\bar{0},\tilde{x}_n) .
\eqno(1.11)
$$
\endproclaim

\smallskip

A contemplation will convince the reader that
the meaning of the summation parameter $n$
in the last expression consists in counting appropriate winding numbers 
of geodesics in $\Cal{X}$ starting at $x$  {\it along}
the closed geodesic $\gamma_0$.

\smallskip

One can similarly write a more informative formula calculating the
whole determinant of the Dirac operator (1.9) 
which involves winding  numbers {\it around} $\gamma_0$.
using the formula (1.6) which
 provides
the phases of cross-ratios in terms of angles and parallel
translations of the relevant geodesic
configurations. We leave this as an exercise for the reader.

\medskip

{\bf 1.4. Genus $\ge 2$ case and Krasnov's Euclidean black holes.}
The construction of the BTZ black hole with Lorentzian signature can
be generalized to other asymptotically $AdS_{2+1}$ solutions, by
prescribing global identifications on the $t=0$ slice $\bold{H}^2$ of
$AdS_{2+1}$, obtained by the action of a discrete subgroup of
$PSL(2,\bold{R})$. Solutions of this type are described in [ABBHP].
They admit a Euclidean version which is a global quotient of
$\bold{H}^3$ by the action of a discrete group of isometries $\Gamma$.
We are especially interested in the case where $\Gamma\subset
PSL(2,\bold{C})$ is a geometrically finite Schottky group. Such
solutions were studied by Krasnov [Kr1], [Kr4], so we refer to them as 
Krasnov black holes. For this class of space--times, in the Euclidean
case, the bulk space is a hyperbolic handlebody of genus $g\geq 2$,
and the surface at infinity is a compact Riemann surface of genus $g$,
with the complex structure determined by the Schottky uniformization.

\medskip

{\bf 1.4.1. Schottky groups and handlebodies.} {\it (i) Loxodromic
elements.} As in 1.2, we choose a 2--dimensional complex
vector space $V$ and study the group $PGL(2)$ and various
spaces upon which it acts. A loxodromic element $g\in PGL(2,V)$,
by definition,  has two different fixed points in $P(V)=\bold{P}^1(\bold{C})$,
the attracting one $z^+(g)$ and the repelling one $z^-(g)$.
The eigenvalue $q(g)$ of $g$ on the complex tangent space to $z^+(g)$
is called the multiplier of $g$. We have $|q(g)|<1.$ 

\smallskip

{\it (ii) Schottky groups.} A Schottky group 
is a finitely generated discrete subgroup  $\Gamma\subset PGL(V)$
consisting of loxodromic 
elements and identity. It is always free; its minimal
number of generators $p$ is called genus. Each Schottky
group of genus $p$ admits a marking. By definition,
this is a family of $2p$ open connected domains $D_1,\dots ,D_{2p}$
in $P(V)$ and a family of generators $g_1,\dots ,g_p\in \Gamma$
with the following properties. The boundary $C_i$ of $D_i$ is a Jordan
curve homeomorphic to $S^1$, closures of $D_i$ are pairwise
disjoint; moreover, $g_k(C_k)\subset C_{p+k}$,
and $g_k(D_k)\subset P(V)\setminus D_{p+k}.$ A marking is called classical,
if all $D_i$ are circles. Every Schottky group
admits a marking, but there are groups for which no classical
marking exists.

\smallskip

{\it (iii) $\Gamma$--invariant sets and their quotients.}
Any Schottky group  $\Gamma$ of genus $p$ acts on $\bold{H}_V$ faithfully
and discretely. The quotient $\Cal{X}_{\Gamma}:=\Gamma\subset \bold{H}_V$
is (the interior of) a handlebody of genus $p$.

\smallskip

Choose a marking and put 
$$
X_{0,\Gamma}:=P(V)\,\setminus\,\cup_{k=1}^p
(D_k\cup\overline{D}_{k+p}),\ \Omega_{\Gamma}:= 
\cup_{g\in\Gamma}\,g(X_{0,\Gamma}).
$$
$\Gamma$ acts on $\Omega_{\Gamma}$ faithfully and discretely,
$X_{0,\Gamma}$ is a fundamental domain for this action, and the quotient
$\Gamma\setminus \Omega_{\Gamma}$ is a complex Riemann surface
of genus $p$. Every Riemann surface admits infinitely many
different Schottky covers.

\smallskip

In the representation above, $\Gamma$ acts upon $\Omega_{\Gamma}$
as on the boundary of a tubular neighborhood of a Cayley graph
of $\Gamma$ associated with generators $g_k$. Since they are free,
the Cayley graph is an infinite tree each vertex of which has
multiplicity $2p$: 
cf. Fig. 4 illustrating this for the case $p=2.$

\bigskip

\input epsf
\midinsert
$$\centerline{\hbox{\epsffile{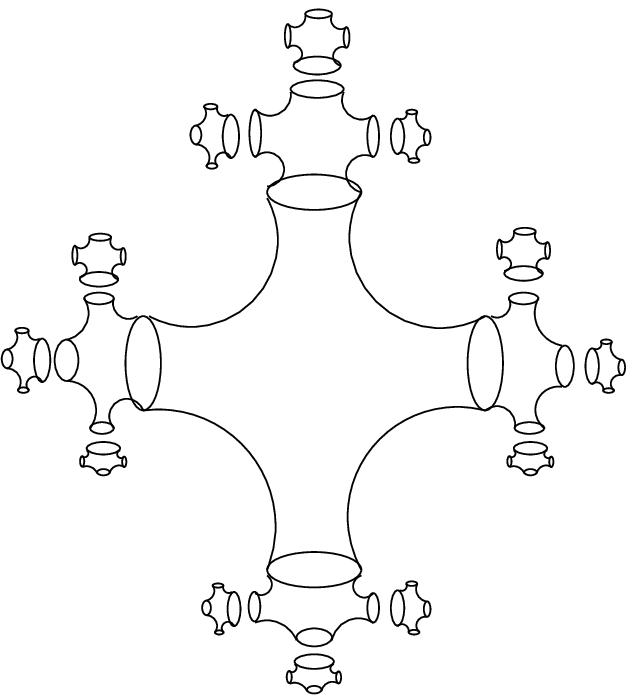}}}$$
\endinsert

\centerline{Fig. 4}

\pagebreak

As above, $X_{\Gamma}$ can be identified with the boundary at infinity
of $\Cal{X}_{\Gamma}$: the set of equivalence classes of
unbounded ends of geodesics in $\Cal{X}_{\Gamma}$ modulo the relation
``distance $=0$.''

\smallskip

A marking of $\Gamma$ induces a marking of the 1--homology
group $H_1(X_{\Gamma},\bold{Z})$. Concretely, denote by
$a_k$ the class of the image of $C_{p+k}$ (with its natural orientation.)
Choose some points $x_k\in C_k$, $k=1,\dots ,p,$ and
pairwise disjoint oriented paths from $x_k$ to $g_k(x_k)$ lying
in $X_{0,\Gamma}.$ Denote by $b_k$ their classes in
$H_1(X_{\Gamma},\bold{Z})$. Clearly, $\{a_k,b_l\}$ form a basis
of this group, satisfying $(a_k,a_l)=(b_k,b_l)=0$,
$(a_k,b_l)=\delta_{kl}.$ Moreover, $a_k$ generate the kernel
of the map $H_1(X_{\Gamma},\bold{Z})\to H_1(\overline{\Cal{X}}_{\Gamma},\bold{Z})$ induced by the inclusion of the boundary.

\smallskip

The complement $\Lambda_{\Gamma}:=P(V)\,\setminus\,\Omega_{\Gamma}$
is the minimal non--empty $\Gamma$--invariant set.
Equivalently, it is the closure of the set of all fixed
points $z^{\pm}(g)$, $g\in \Gamma ,g\ne \roman{id}$,
or else the set of limit points of any orbit $\Gamma z_0$,
$z_0\in \bold{H}_V \cup P(V).$ 

\smallskip

If $g=1$, $\Lambda_{\Gamma}$ consists of two points which
can be chosen as $0,\infty.$ For $g\ge 2$, $\Lambda_{\Gamma}$
generally is
an uncountable   Cantor set (fractal). This is the main source of complications 
(and interesting developments).
Denote by $a(\Gamma )$ the Hausdorff dimension of $\Lambda (\Gamma ).$
It can be characterized as the convergence abscissa
of any Poincar\'e series
$$
\sum_{g\in\Gamma}\left|\frac{dg(z)}{dz}\right|^s
$$
where $z$ is any coordinate function on $P(V)$
with a zero and a pole in $\Omega_{\Gamma}.$
Generally $0< a(\Gamma )<2.$
Convergence of our holography formulas below
will hold only for $a(\Gamma )<1.$ For other characterizations of $a(\Gamma )$,
see [Man2], p. 236, and the references therein.

\smallskip

Geodesics in the bulk space
$\bold{H_V}$ with ends on $\Lambda_{\Gamma}$ become exactly all
bounded geodesics in the quotient $\Cal{X}_{\Gamma}$.
Their convex hull $\Cal{C}_{\Gamma}$ is called the convex core
of $\Cal{X}_{\Gamma}.$ The group $\Gamma$ is
geometrically finite if the convex core ${\Cal C}_\Gamma$ is of finite
volume. In this case, the core ${\Cal C}_\Gamma$ is a compact
3--manifold with boundary, which is homeomorphic to 
and a strong deformation retract of ${\Cal X}_\Gamma$.

\medskip

{\bf 1.4.2. AdS and Euclidean black holes.} Consider a Fuchsian
Schottky group $\Gamma$ acting on $\bold{H}^2$. The resulting quotient
space is a non--compact Riemann surface with a certain number of
infinite ends. The genus of the surface and the number of ends depend
on the Schottky group, for instance, both topologies shown in
the Fig. 5 arise as quotients of $\bold{H}^2$ by a Schottky group
with two generators.

\pagebreak

\input epsf
\midinsert
$$\centerline{\hbox{\epsffile{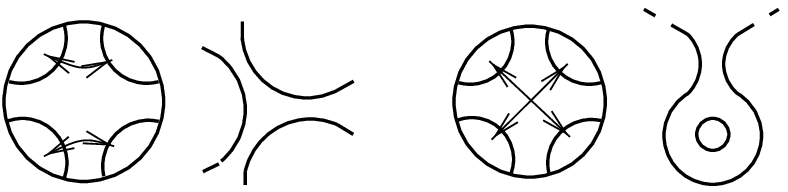}}}$$
\endinsert
\centerline{Fig. 5} 

\medskip

An asymptotically AdS non--spinning black hole is obtained by
extending these identifications globally to $AdS_{2+1}$, or, in other
words, by evolving the $t=0$ slice forward and backward in time.
The geodesic surfaces extending the geodesics in the boundary of the
fundamental domain in the $t=0$ slice develop singularities in both
forward and backward direction (see [ABBHP], [Kr1]) as illustrated in
Fig. 6. 

\bigskip

\input epsf
\midinsert
$$\centerline{\hbox{\epsffile{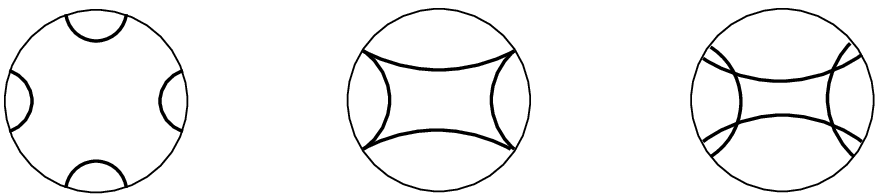}}}$$
\endinsert

\centerline{Fig. 6}

\medskip

The procedure used by Krasnov [Kr1] to construct the Euclidean version of
these black holes follows the same line as in the case of the BTZ
black hole, namely, the $t=0$ slice is identified with a hyperplane in
$\bold{H}^3$ and the geodesics in this hyperplane are continued to
geodesic surfaces in $\bold{H}^3$. The resulting quotients are special
cases (non--rotating black holes) of the handlebodies ${\Cal
X}_\Gamma$ constructed above in 1.4.1, in the case of real Schottky
parameters. The general case of 1.4.1 includes also the more general
case of spinning black holes considered by Krasnov in [Kr4].

\smallskip

Since Fuchsian Schottky groups are classical Schottky groups,
the black holes obtained by the construction of Krasnov as Euclidean
versions of the AdS black holes of [ABBHP] are quotients of
$\bold{H}^3$ by a classical Schottky group on $p$ generators, and the 
fundamental domain is a region in $\bold{H}^3$ delimited by $2p$ 
pairwise disjoint geodesic half spheres. 

\smallskip

As observed in [BKSW], the kinematic part of the Maldacena
correspondence for spacetimes that are global quotients of
$\bold{H}^3$ by a geometrically finite discrete group of isometries is
provided by the correspondence between hyperbolic structures on the
bulk space and conformal structures on the boundary at infinity,
[Sul]. (cf.~ also [Kh] on the correspondence between hyperbolic and
conformal geometry viewed in the light of holography.)
Below we will complement this by providing some 
dynamical content for the case of the Krasnov black holes. 

\medskip

{\bf 1.5. Abelian differentials and Green functions
on Schottky covers.} In this subsection, we will calculate
Green's functions of the form (1.2) for curves with a Schottky
cover. The differentials of the third kind which can be obtained
by a direct averaging of simple functions do not necessarily
have pure imaginary periods. To remedy this, we will have
to subtract from them some differentials of the first kind.
Therefore we will start with the latter.

\medskip

{\bf 1.5.1. Differentials of the first kind.} In the genus one case,
if $z$ is the projective coordinate whose divisor consists
of the attractive and repelling point of a generator of $\Gamma$,
a differential of the first kind can be written as
$$
\omega = d\,\roman{log}\,z=
d\,\roman{log}\,\frac{w_{(0)-(\infty )}(z)}{w_{(0)-(\infty )}(z_0)}=
d\,\roman{log}\,\langle 0,\infty ,z,z_0\rangle
$$
where $z_0$ is any point $\ne 0,\infty $.
Generally, an appropriate averaging of this formula produces 
a differential of the first kind $\omega_g$ for any $g\in \Gamma .$
In the following we assume that a marking of $\Gamma$ is chosen.
Denote by $C(|g)$ a set of representatives of
$\Gamma /(g^{\bold{Z}})$, by $C(h|g)$ a similar set for
$(h^{\bold{Z}})\setminus\Gamma /(g^{\bold{Z}})$,
and by $S(g)$ the conjugacy class
of $g$ in $\Gamma$. Then we have for any $z_0\in \Omega_{\Gamma}$:

\medskip

\proclaim{\quad 1.5.2. Proposition} (a) If $a(\Gamma )<1,$
the following series converges absolutely for $z\in\Omega_{\Gamma}$
and determines (the lift to $\Omega_{\Gamma}$ of) a differential of
the first kind on $X_{\Gamma}$: 
$$
\omega_g=\sum_{h\in C(|g)}\, d_z\,\roman{log}\,
\langle hz^+(g), hz^-(g),z,z_0\rangle\, .
\eqno(1.12)
$$
This differential does not depend on $z_0$, and depends on $g$ additively.

\smallskip

If the class of $g$ is primitive (i.~e. non--divisible in $H$),
$\omega_g$ can be rewritten as 
$$
\omega_g=\sum_{h\in S(g)}\, d_z\,\roman{log}\,
\langle z^+(h), z^-(h),z,z_0\rangle\, .
\eqno(1.13)
$$
\smallskip

(b) If $g_k$ form a part of the marking of $\Gamma$, and $a_k$
are the homology classes described in 1.4.1 (iii), we have
$$
\int_{a_k}\omega_{g_l} =2\pi i\,\delta_{kl}.
\eqno(1.14)
$$
It follows that the map $g\,\roman{mod}\,[\Gamma ,\Gamma ] \mapsto \omega_g$
embeds $H:=\Gamma/[\Gamma ,\Gamma ]$ as a sublattice in the space of
all differentials 
of the first kind.

\smallskip

(c) Denote by $\{b_l\}$ the complementary set of
homology classes in $H_1(X_{\Gamma},\bold{Z})$ as in 1.4.1.
Then we have for $k\ne l$, with an appropriate choice
of logarithm branches:
$$
\tau_{kl}:=\int_{b_k}\omega_{g_l}=\sum_{h\in C(g_k|g_l)}
\roman{log}\,\langle z^+(g_k), z^-(g_k), hz^+(g_l), hz^-(g_l)\rangle\, .
\eqno(1.15)
$$
Finally
$$
\tau_{kk}= \roman{log}\,q(g_k)+\sum_{h\in C_0(g_k|g_k)}
\roman{log}\,\langle z^+(g_k), z^-(g_k), hz^+(g_k), hz^-(g_k)\rangle\, .
\eqno(1.16)
$$
where in $C_0(g_k|g_k)$ is $C(g_k|g_k)$ without the identity class.
\endproclaim

\smallskip

For proofs, see [Man2], \S 8, and [ManD]. Notice that our notation
here slightly differs from [Man2]; in particular,
$\tau_{kl}$ here corresponds to $2\pi i\tau_{kl}$ of [Man2].

\smallskip

In the holography formulas below we will use (1.15) and (1.16)
in order to calculate $\roman{Re}\,\tau_{kl}.$ The ambiguity
of phases can then be discarded, and the cross--ratios 
must be replaced by their absolute values.
Each resulting term can then be interpreted via a configuration of geodesics
in the bulk spaces  $\bold{H}^3$ and $\Cal{X}_{\Gamma}$, similar to those
displayed in Fig. 3a and Fig. 3b.

\medskip

{\bf 1.5.3. Differentials of the third kind and Green's functions.}
Let now $a,b\in \Omega_{\Gamma}.$ Again assuming $a(\Gamma )<1,$
we see that the series 
$$
\nu_{(a)-(b)}:=\sum_{h\in\Gamma} d_z \roman{log}\,
\langle a,b,hz,hz_0\rangle
\eqno(1.17)
$$
absolutely converges and represents the lift to $\Omega_{\Gamma}$
of a differential of the third kind with residues
$\pm 1$ at the images of $a,b.$ Moreover, its $a_k$ periods vanish.
Therefore, any linear combination $\nu_{(a)-(b)}-\sum_l X_l(a,b)\omega_{g_l}$
with real coefficients $X_l$ will have pure imaginary
$a_k$--periods in view of (1.14). If we find $X_l$ so that
the real parts of the $b_k$--periods of
$\omega_{(a)-(b)}:=\nu_{(a)-(b)}-\sum_l X_l(a,b)\omega_{g_l}$ 
vanish, we will be able to use this differential
in order to calculate conformally invariant Green's functions.
Hence our final formulas look as follows. 

\smallskip

Equations for
calculating $X_l(a,b)$:
$$
\sum_{l=1}^p X_l(a,b)\,\roman{Re}\,\tau_{kl}=
\roman{Re}\,\int_{b_k}\nu_{(a)-(b)}=
\sum_{h\in S(g_k)} \roman{log}\,|\langle
a,b,z^+(h),z^-(h)\rangle |\, .
\eqno(1.18)
$$
Here $k$ runs over $1,\dots ,p,$ $\roman{Re}\,\tau_{kl}$ are
calculated by means of (1.15) and (1.16), 
and $b_k$--periods of $\nu_{(a)-(b)}$ are given in \S 8 of [Man2].

\smallskip

Moreover,
$$
\roman{Re}\,\int_d^c \nu_{(a)-(b)}=\sum_{h\in\Gamma}
\roman{log}\,|\langle
a,b,hc,hd\rangle |\, ,
\eqno(1.19)
$$
$$
\roman{Re}\,\int_d^c \omega_{g_l}=
\sum_{h\in S(g_l)}
\roman{log}\,|\langle
z^+(h),z^-(h), c,d\rangle |\, ,
\eqno(1.20)
$$
Hence finally
$$
g((a)-(b),(c)-(d))=
$$
$$
\sum_{h\in\Gamma}
\roman{log}\,|\langle
a,b,hc,hd\rangle |- \sum_{l=1}^p X_l(a,b)\,
\sum_{h\in S(g_l)}
\roman{log}\,|\langle
z^+(h),z^-(h), c,d\rangle |\, .
\eqno(1.20)
$$
Here we have to thank Annette Werner for 
correcting the last formula in [Man2].

\medskip

{\bf 1.6. Discussion.} (i) The most straightforward way
to interpret formulas (1.3), (1.4), (1.11), (1.20)
is to appeal to the picture
of holographic particle detection of [BR]. In this picture,
Green functions on the boundary detect geodesic movement
and collisions of massive particles in the bulk space.
Particles, being local objects, exist in the semiclassical limit.

\smallskip

More precisely, consider
in the bulk space the theory of a scalar field of mass $m$.
The propagator, in the  notation of [BR] p.7, is
$$ 
G(B(z),B(-z))= \int {\Cal D}P e^{i\Delta \ell(P)}, 
\eqno(1.21) 
$$
where $\ell(P)$ is the length of the path $P$,$\Delta=1+\sqrt{1+m^2}$,
and the points $B(\pm z)$ in the bulk space correspond to some
parameterized curve $b(\pm z)$ on the boundary at infinity, in the
sense that the $B(\pm z)$ lie on a hypersurface obtained by introducing a
cutoff on the bulk space.

\smallskip

In the semiclassical WKB approximation, the right hand side of (1.21)
localizes at the critical points of action. Thus, it becomes a sum over
geodesics connecting the points $B(\pm z)$,
$$ 
G(B(z),B(-z))= \sum_\gamma e^{-\Delta \ell(\gamma)}. 
\eqno(1.22) 
$$
This has a logarithmic divergence when the cutoff $\epsilon \to 0$,
that is, when the points $B(\pm z)$ approach the corresponding points
on the boundary at infinity. 

\smallskip

On the other hand, for the CFT on the boundary (in the case where the
bulk space is just $AdS_3$),  the boundary propagator 
is taken in the form in the form (pp. 6--7
of [BR]) 
$$  
\langle O(x), O(x') \rangle = \frac{1}{|x-x'|^{2\Delta}}. 
$$
In the case where the bulk space is globally $AdS_3$, there is
an identification of the propagators as the cutoff parameter
$\epsilon \to 0$ 
$$ 
T(z) \equiv \log G(B(z),B(-z)), 
$$
after removing the logarithmic divergence, where
$$ 
T(z) =\log \langle O(b(z)), O(b(-z)) \rangle. 
$$
The appearance of the geodesic propagator
(1.21) in the bulk space, written in the form (1.22) is somewhat
similar to our exact formulas written in terms of geodesic
configurations.

\smallskip

Moreover, 
passing to the Euclidean case, 
and reading our formula (1.3) for the genus zero case in
this context provides a neater way of identifying propagators
on bulk and 
boundary which does not require any cutoff. For assigned points on the
boundary $\bold{P}^1(\bold{C})$, instead of choosing corresponding 
points in the bulk space $B(\pm z)$ with the help of a cutoff
function and then comparing propagators in the limit, any choice of a
divisor $(a)-(b)$ determines the points in the bulk space
$a*\{c,d \}$ and $b*\{ c,d \}$ in $\bold{H}^3$, for boundary points
$c,d \in \bold{P}^1(\bold{C})$, and a corresponding exact identification of
the propagators.

\smallskip

If we then let $a\to c$ and $b\to d$, in (1.3) both the Green function
and the geodesic length have a logarithmic divergence, as the points
$a*\{c,d \}$ and $b*\{ c,d \}$ also tend to the boundary points $c$
and $d$, and this recovers the identification of the propagators used
by the physicists as a limit case of formula (1.3), without any need to
introduce cutoff functions.

\smallskip

Notice, moreover, that the procedure of \S 1.5.3, and in particular
our (1.18) to compute the coefficients $X_l(a,b)$ is analogous to the
derivation of the bosonic field propagator for algebraic curves in
[FeSo], with the sole difference that, in the linear combination
$$ \omega_{(a)-(b)}:=\nu_{(a)-(b)}-\sum_l X_l(a,b)\omega_{g_l}, $$
cf.~equation (3.6) of [FeSo], the differentials of the third kind
$\nu_{(a)-(b)}$ are determined in our (1.17) by the data of the
Schottky uniformization, while, in the case considered in [FeSo], they
are obtained by describing the algebraic curve as a branched cover of
$\bold{P}^1(\bold{C})$. Then our (1.18) corresponds to (3.9) of
[FeSo], and Proposition 1.5.2 shows that the bosonic field  propagator
on the algebraic curve $X(\bold{C})$, described by the Green function,
can be expressed in terms of geodesics in the bulk space.

\medskip 

(ii)  K.~Krasnov in [Kr1] (cf. also [Kr2]--[Kr4]) 
establishes another holography
correspondence which involves CFT interpreted as geometry of the
Teichm\"uller or Schottky moduli space rather than that
of an individual Riemann surface and its bulk handlebody.
In his picture, the relevant CFT theory is
the Liouville theory (existence of which is not
yet fully established). An appropriate action for Liouville theory
in terms of the Schottky uniformization was
suggested  L.~Takhtajan and P.~Zograf in [TaZo]. Krasnov
identifies the value of this action at the stationary (``uniformizing'')
point with the regularized volume of the respective Euclidean bulk space.
According to [TaZo], this value provides
the K\"ahler potential for the Weil--Petersson metric
on the moduli space.

\smallskip

It would be interesting to clarify the geometric meaning
of Krasnov's regularized volume. Can it be calculated
through the volume of the convex core of the bulk space?
In the genus one case the answer is positive:
both quantities are proportional to the length
of the closed geodesic.

\smallskip

A recent preprint of J.~Brock [Br1] establishes an
approximate relationship
between the Weil--Petersson metric and volumes of
convex cores in a different, but related situation.
Namely, instead of giving a local formula
for the WP--distance ``at a point'' $X$, it provides
an approximate formula for this distance 
between two Riemann surfaces $X,Y$ which are far apart.
The handlebody $\Cal{X}$ filling $X$ is replaced by the
quasi--Fuchsian hyperbolic 3--manifold
$Q(X,Y)$ arising in the Bers simultaneous uniformization
picture ([Be]) and having $X\cup Y$
as its conformal boundary at infinity.
It turns out that at large distances $\ell_{WP}(X,Y)$
is comparable with the volume of $core\,Q(X,Y).$

\smallskip

We expect that an exact formula relating these two quantities
exists and might be derived using a version
of Krasnov's arguments. 

In fact, the Krasnov black holes also have a description in terms
of Bers simultaneous uniformization. By the results of Bowen [Bow],
a collection $C_0$ of pairwise disjoint rectificable arcs in
$X_{0,\Gamma}$ with ends at $x_k\in C_k$ and $g_k(x_k)$, as described
in \S 1.4.1(iii), determine a quasi--circle $C = \cup_{\gamma \in
\Gamma} \gamma C_0$. The quotient $(C\cap \Omega_\Gamma)/\Gamma$ consists
of a collection of closed curves in $X_\Gamma$ whose homology classes
give the $b_k$ of \S 1.4.1(iii). The quasi--circle $C$ divides
$\bold{P}^1(\bold{C})$ into two domains of Bers simultaneous
uniformization, with the handlebody ${\Cal X}_\Gamma$ (topologically a
product of a non--compact Riemann surface and an interval) in the
role of $Q(X,Y)$.
This fits in with the results of Krasnov on the generally rotating
case of Krasnov black holes discussed in [Kr4]. 

\pagebreak

{\bf 1.7. Non--archimedean holography.} According to various speculations, 
space--time at the Planck scale should 
be enriched with non--archimedean geometry, possibly in ad\`elic form,
so that space--time can be seen simultaneously at all non--archimedean and
archimedean places. From this perspective, it is worth observing that
the holography correspondence described in \S 1 admits a natural
extension to the non--archimedean setting. In fact, the results of
[Man2] on the Green functions on Riemann surfaces with Schottky
uniformization and configurations of geodesics in the bulk space were
motivated
by the theory of $p$--adic Schottky groups and Mumford curves: cf.~[Mum],
[ManD], [GvP].

\smallskip

In the non--archimedean setting, we consider a finite extension $K$ of
$\bold{Q}_p$. Anti de Sitter space, or rather its
Euclidean analog $\bold{H}^3$, is replaced by the Bruhat--Tits tree
${\Cal T}$ with the set of vertices 
$$ 
{\Cal T}^0= \{ \text{$A$--lattices of rank 2 in a 2--dim
$K$--space $V$} \}/ K^* 
\eqno(1.22) 
$$ 
where $A$ is the ring of integers of $K$. Vertices have valence 
$|\bold{P}^1(A/\bold{m})|$, where $\bold{m}$ is the maximal ideal, and
the length of each edge connecting two nearby vertices is $\log\,|
A/\bold{m}|$. The set of ends of the tree ${\Cal T}$ can be identified
with $\bold{P}^1(K)$: this is the analog of the conformal boundary. 

\smallskip 

The analog in [ManD] of the formulas of Lemma 1.1.3,
gives a
quantitative formulation of the holographic correspondence
in this non--archimedean setting, with the basic Green function on
${\Cal T}$ given by
$$ 
G_{\mu(u)} (x,y) = \roman{dist}_{\Cal T} (u, \{ x,y \}), 
\eqno(1.23) 
$$
where the metric on the Bruhat--Tits tree ${\Cal T}$ is defined by
assigning the length $\log\,|A/\bold{m}|$ to each edge, so that
(1.23) computes the length of the shortest chain of edges
connecting the vertex $u$ to the doubly infinite path in the tree
containing the vertices $x,y$.

\smallskip 
 
A triple of points in
$\bold{P}^1(K)$ determines a unique vertex  $v\in {\Cal T}^0$ where the
three ends connecting $v$ to the given points in $\bold{P}^1(K)$ start
along different edges. This configuration of edges is called
a ``cross--roads'' in [Man2]. It provides an analog of the Feynman diagram
of \S 2.4 of [Wi], where currents are inserted
at points on the boundary and the interaction takes place in the
interior, with half infinite paths in the Bruhat--Tits tree acting as
the gluon propagators. Such propagators admit a nice arithmetic
description in terms of reduction modulo the 
maximal ideal $\bold{m}$. 

\smallskip

For a subgraph of ${\Cal T}$ given by the half infinite path
starting at a given vertex $v\in {\Cal T}^0$ with end $x\in
\bold{P}^1(K)$, let $\{ v_0=v, v_1, \ldots, v_n \ldots \}$ be the
sequence of vertices along this path. We can define a
`non--archimedean gluon propagator' as such a graph together with  
the maps that assign to each finite path
$\{ v_0, \ldots ,v_n \}$ the reduction of $x$ modulo
$\bold{m}^n$.

\smallskip

Consider the example of the elliptic curve with the Jacobi--Tate
uniformization $K^* /(q^\bold{Z})$, with $q\in K^*$, $|q|<1$.
The group $q^\bold{Z}$ acts on ${\Cal T}$ like the cyclic group
generated by an arbitrary hyperbolic element $\gamma \in
PGL(2,K)$. The unique doubly infinite path in ${\Cal T}$ with
ends at the pair of fixed points $x^\pm$ of $\gamma$ in
$\bold{P}^1(K)$ gives rise to a closed ring in the quotient ${\Cal T}
/\Gamma$.  

\smallskip 

The quotient space ${\Cal T} /\Gamma$ is the non--archimedean version
of the BTZ black hole, and this closed ring is the event horizon. From
the vertices of this closed ring 
infinite ends depart, which correspond to the reduction map $X(K)\to
X(A/\bold{m})$.

\smallskip

Subgraphs of the graph ${\Cal T} /\Gamma$ correspond to
all possible Feynman diagrams of propagation between boundary sources
on the Tate elliptic curve $X(K)$ and interior vertices on the closed
ring.  

\smallskip

In the case of higher genus, the Schottky group $\Gamma$ is a purely
loxodromic free discrete subgroup of $PSL(2,K)$ of rank $g\ge 2$. The
doubly infinite 
paths in ${\Cal T}$ with ends at the pairs of fixed points
$x^\pm(\gamma)$ of the elements $\gamma\in \Gamma$ realize ${\Cal
T}_\Gamma$ as a subtree of ${\Cal T}$. This is the analog of
realizing the union of fundamental domains $\cup_\gamma
\gamma({\Cal F})$ as a tubular neighborhood of the Cayley graph of
$\Gamma$ in the archimedeam case.
The ends of the subtree ${\Cal T}_\Gamma$ constitute the limit set
$\Lambda_\Gamma \subset \bold{P}^1(K)$. The 
complement $\Omega_\Gamma = \bold{P}^1(K)\smallsetminus \Lambda_\Gamma$
gives the uniformization of the Mumford curve $X(K)\simeq
\Omega_\Gamma /\Gamma$. This, in turn, can be identified with
the ends of the quotient graph ${\Cal T}/\Gamma$. 

\smallskip 

The quotients ${\Cal X}_\Gamma={\Cal T}/\Gamma$ are 
non--archimedean Krasnov black holes, with boundary
at infinity the Mumford curve $X(K)$. Currents at points in $X(K)$
propagate along the half infinite paths in the black hole that reaches 
a vertex on ${\Cal T}_\Gamma/\Gamma$. Propagation between interior
points happen along edges of ${\Cal T}_\Gamma/\Gamma$, and loops in this
graph give rise to quantum corrections to the correlation functions of
currents in the boundary field theory, as happens with the Feynman
diagrams of [Wi].

\medskip

{\bf 1.8. Holography and arithmetic topology.} We have seen that, for
an arithmetic surface $X_{\bold{Z}} \to \text{Spec}\, \bold{Z}$, it is
possible to relate the geometry at arithmetic infinity to the physical
principle of holography. Over a prime $p$, in the case of curves with
maximally degenerate reduction, it is also possible to interpret the
resulting Mumford theory of $p$--adic Schottky uniformization in terms
of an arithmetic version of the holography principle. One can
therefore formulate the question of whether some other arithmetic
analog of holography persists for closed fibers $X_{\bold{Z}} \mod
p$. 

\smallskip

A very different picture of the connection between 3--manifolds
and arithmetics exists in the context of {\it arithmetic topology}, a
term introduced by Reznikov [Rez] to characterize a dictionary of
analogies between number fields and 3--manifolds. See also a nice
overview by McMullen [Mc]. 

\smallskip

According to this dictionary, if $L$ is a number field and ${\Cal O}_L$
its ring of algebraic integers, then $B=\text{Spec}\, {\Cal O}_L$ is an
analog of a 3--manifold, with primes representing loops (knots in
a 3--manifold). In our case, with $B=\text{Spec}\, \bold{Z}$, the local
fundametal group $Gal(\bar\bold{F}_p/\bold{F}_p) \cong \hat \bold{Z}$
is generated by the Frobenius $\sigma_p : x \mapsto x^p$ acting 
on $\bar\bold{F}_p$. 

\smallskip

The fiber of $X$ over a prime $p$, in the dictionary of
arithmetic topology, may be regarded as a 3--manifold that fibers
over a circle. In fact, for a fixed prime $\ell$, let $S$ be the union
of $\ell$ and the set of primes where $X$ has bad reduction. Let $B=
\text{Spec}\, S^{-1} \bold{Z}$. This satisfies
$\hat\pi_1(B)=Gal(\bar\bold{Q}_S/\bold{Q})$. For $p\notin S$, the
$\ell$--adic Galois representation  
$$ \hat\rho_\ell : Gal(\bar\bold{Q}_S/\bold{Q}) \to Aut
H_1(X,\bold{Z}_\ell)=GL(2g, \bold{Z}_\ell) $$
gives an arithmetic version of the monodromy, see [Mc], with the
Frobenius $\sigma_p$ that lifts to an element of
$Gal(\bar\bold{Q}_S/\bold{Q})$.  
In the arithmetic topology dictionary, a prime $p$ corresponds to a
``loop'' in the ``3--manifold'' $B$, hence the fiber $X_{\bold Z} \mod
p$ together with the Frobenius element $\sigma_p$ can be regarded as
the data of a 3--manifold that fibers over the ``circle'' $p$.

\smallskip

The question of a holographic correspondence for these arithmetic
analogs of mapping tori may be related to results of J.Brock [Br2] on 
3--manifolds that fiber over the circle, where the hyperbolic volume is
related to the translation length of the monodromy, in the same way
that relates the hyperbolic volume of the convex core to the
Weil--Petersson distance of the surfaces at infinity in the case of
Bers' simultaneous uniformization in the main result of [Br1]. 

\smallskip

In our perspective, this result of [Br2] can be regarded as an
extension of a form of holographic correspondence from the
case of hyperbolic 3--manifolds with infinite ends and asymptotic
boundary surfaces, to the case of a compact hyperbolic 3--manifolds which
fibers over the circle, with the information previously carried by the
boundary at infinity now residing in the fiber and 
monodromy. Thus, it is possible to ask whether, under the dictionary
of arithmetic topology, a similar form of holographic correspondence
exists for the fibers $X_{\bold Z} \mod p$ regarded as arithmetic
analogs of a 3--manifold fibering over the circle with monodromy
$\sigma_p$. It is possible that such correspondence may be related to
another analogy of arithmetic topology, which interprets the quantity
$|Tr (\sigma_p)|$ as a measure of the ``hyperbolic length''  
of the loop representing the prime $p$ (cf.~[Mc] Remark on p.134).

\bigskip

\centerline{\bf \S 2. Modular curves as holograms}

\medskip

In this section we suggest a different type of holography 
correspondence, this time
related to $AdS_{1+1}$ and its Euclidean version $\bold{H}^2$.

\smallskip

In the case we consider, the bulk spaces will be modular curves. They
are global quotients of the hyperbolic plane $\bold{H}^2$  by a finite
index subgroup $G$ of $PSL(2,\bold{Z})$. We identify
$\bold{H}^2$ with the upper complex half--plane endowed
with the metric $dx^2/y^2$ of curvature $-1$. Its boundary at infinity
is then $\bold{P}^1(\bold{R})$.

\smallskip

Modular
curves have a very rich arithmetic structure, forming the essential part
of the moduli stack of elliptic curves. In this
classical setting, the modular curves have a natural
algebro--geometric compactification, which consists of adding finitely
many points at infinity, the cusps $G\backslash
\bold{P}^1(\bold{Q})$. Cusps are the only boundary points at which 
$G$ acts discretely (with stabilizers of finite index).
The remaining part of the conformal boundary (after factorization)
is not visible
in algebraic (or for that matter analytic or $C^{\infty}$) geometry, 
because irrational orbits of $G$ in $\bold{P}^1(\bold{R})$
are dense.

\smallskip

In [ManMar] and [Mar] some aspects of the classical
geometry and arithmetics of modular curves, such as modular symbols,
the modular complex, and certain classes of modular forms, are
recovered in terms of the non--commutative 
boundary $G\backslash \bold{P}^1(\bold{R})$
which is a non--commutative space in the
sense of Connes, that is, a $C^*$--algebra Morita equivalent to the
crossed product of  $G$ acting on some function ring of
$\bold{P}^1(\bold{R})$. This 
way, the full geometric boundary of $\bold{P}^1(\bold{R})$ of
$\bold{H}^2$ is considered as part of the compactification, instead of
just $\bold{P}^1(\bold{Q})$.  
We argue here that this is the right notion of boundary to consider in
order to have a holography correspondence for this class of bulk
spaces. In particular, since we strive to establish that
the bulk spaces and their boundaries carry essentially the same
information, we call the quotients  $G\backslash \bold{P}^1(\bold{R})$
{\it non--commutative modular curves.}

\medskip

{\bf 2.1. Non--commutative modular curves.}
In the following $\Gamma = PSL(2,\bold{Z})$
and $G$ is a finite index subgroup of $\Gamma$.
Denoting by $\bold{P}$ the coset space
$\bold{P}=\Gamma/G$, we can represent the modular 
curve $X_G:=G\backslash \bold{H}^2$ as the
quotient
$$ 
X_G= \Gamma \backslash (\bold{H}^2 \times \bold{P}), 
\eqno(2.1) 
$$
and its non--commutative boundary as the  $C^*$--algebra
$$ 
C(\bold{P}^1(\bold{R})\times \bold{P}) \rtimes \Gamma 
\eqno(2.2) 
$$
Morita equivalent to $C(\bold{P}^1(\bold{R}))\rtimes G$.

\smallskip

There is a dynamical system associated to the equivalence relation
defined by the action of a Fuchsian group of the first kind on its
limit set, as in the case of our $G\backslash \bold{P}^1(\bold{R})$.
The dynamical system can be described as a Markov map $T_G : S^1 \to
S^1$ as in [BowSer]. 

\smallskip

In [ManMar] we gave a different formulation in terms of a dynamical
system related to the action of $\Gamma$ on
$\bold{P}^1(\bold{R})\times \bold{P}$. This dynamical system
generalizes the classical shift of the continued fraction
expansion in the form
$$ T: [0,1]\times \bold{P} \to [0,1]\times \bold{P} $$
$$ T(x,t) = \left( \frac{1}{x} - \left[ \frac{1}{x} \right],
\left(\matrix -[1/x] & 1 \\ 1 & 0 \endmatrix \right)\cdot t
\right). 
\eqno(2.3) 
$$
Some aspects of the non--commutative geometry at the
boundary of modular curves can be derived from an analysis of the
ergodic theory of this dynamical system, cf.~[ManMar], [Mar].

\medskip

{\bf 2.2. Holography.} The $1+1$--dimensional Anti de Sitter
space--time $AdS_{1+1}$ has $SL(2,\bold{R})$ as group of
isometries. Passing to Euclidean signature, $AdS_{1+1}$ is replaced by
$\bold{H}^2$, so that we can regard the modular curves $X_G$ as
Euclidean versions of space--times obtained as global quotients of
$AdS_{1+1}$ by a discrete subgroup of isometries. Notice that, unlike
the case of spacetimes with $AdS_{2+1}$ geometry, the case of
$AdS_{1+1}$ space--times is relatively little understood, though some
results on $AdS_{1+1}$ holography are formulated in [MMS], [Str]. We
argue that one reason for this is that a picture of holography for
$AdS_{1+1}$ space--times should take into account the possible
presence of non--commutative geometry at the boundary.  

\smallskip

There are three types of results from [ManMar] that can be regarded as
manifestations of the holography principle. On the bulk
space, these results can be formulated in terms of the
Selberg zeta function, of certain classes of modular
forms of weight two, and of modular symbols, respectively.

\medskip

{\bf 2.2.1. Selberg zeta function.}
In order to formulate our
 first results, we consider the Ruelle
transfer operator for the shift (2.3),
$$ 
(L_s f)(x,t)= \sum_{k=1}^{\infty} \frac{1}{(x+k)^{2s}}
f \left(\frac{1}{x+k}, \left( \matrix 0 & 1\\
1 & k \endmatrix \right) \cdot t\right). 
\eqno(2.4) 
$$

On a suitable Banach space of functions (cf.~[May], [ManMar]), the
operator $L_s$ is nuclear of order zero for $Re(s)>1/2$, hence it has
a Fredholm determinant 
$$ 
\det (1- L_s) = \exp \left( - \sum_{\ell =1}^\infty \frac{\text{Tr}
L_s^\ell}{\ell} \right). 
\eqno(2.5) 
$$

The Selberg zeta function for the modular curve $X_G$ encodes the
length spectrum of the geodesic flow. Via the Selberg trace formula,
this function also encodes information on the spectral properties of
the Laplace--Beltrami operator. In terms of closed geodesics, we have
$$ 
Z_G(s)= \prod_{\gamma\in \text{Prim}} \prod_{m=0}^\infty  \left(
1-e^{-(s+m) \,\text{length}(\gamma)} \right), 
\eqno(2.6)
$$
where $\text{Prim}$ is the set of primitive closed geodesics in $X_G$. 
We have the following result [ManMar] (see also [ChMay], [LewZa1],
[LewZa2], [May]). 

\medskip

\proclaim{\quad 2.2.2. Proposition} Consider a finite index subgroup
$G\subset \Gamma$, with $\Gamma= PSL(2,\bold{Z})$ or
$PGL(2,\bold{Z})$. In the case $\Gamma= PGL(2,\bold{Z})$ we have
$$ Z_G(s) = \det (1- L_s), 
\eqno(2.7) 
$$
and in the case $\Gamma= PSL(2,\bold{Z})$ we have
$$ Z_G(s) = \det (1- L_s^2). 
\eqno(2.8) 
$$
\endproclaim

\medskip

We can interpret this statement as an instance of holography
correspondence, if we regard the left hand side of (2.7) and (2.8) as
a partition function on the bulk space, and the right hand side as the
corresponding boundary field theory. More precisely, the results of 
[Lew], [LewZa1], [LewZa2] provide an explicit correspondence between
eigenfunctions of the transfer operator $L_s$ and eigenfunctions of
the Laplacian (Maass wave forms). This explicit transformation
provides a kind of holography correspondence  between fields on the bulk space
and a theory on the boundary, which can be interpreted as a
lattice spin system with the shift operator (2.3). 

\smallskip

To make a connection to the point of view of Arakelov geometry
considered in \S 1, it is known that the Arakelov Green
function evaluated at two different cusps can be estimated in terms of
the constant term of the Laurent expansion around $1$ of the 
logarithmic derivative of the Selberg zeta function, e.g.~ in the
case of $G=\Gamma_0(N)$. This means that, by Proposition
2.2.1, such estimates can be given in terms of the transfer operator
$L_s$, which only depends on the boundary (2.2) of $X_G$.

\medskip

{\bf 2.2.3. Modular symbols.} In the classical theory of modular
curves, modular symbols are the homology classes 
$$ \varphi(s)=\{ g(0), g(i\infty) \}_G \in H_1(X_G,\text{cusps}, \bold{Z})
\eqno(2.9) $$ 
with $gG=s\in \bold{P}$, determined by the image in $X_G$ of geodesics
in $\bold{H}$ with ends at points of $\bold{P}^1(\bold{Q})$. 

\smallskip

In [ManMar] we have shown that the homology $H_1(X_G,
\text{cusps}, \bold{Z})$ can be described canonically in terms of the
boundary (2.2) in the following way.

\medskip

\proclaim{\quad 2.2.4. Proposition} In the case
$\Gamma=PSL(2,\bold{Z})=\bold{Z}/2 *\bold{Z}/3$, the Pimsner six term 
exact sequence ([Pim]) for the $K$--theory of the crossed product
$C^*$--algebra (2.2) gives a map
$$ 
\alpha: K_0(C(\bold{P}^1(\bold{R}) \times \bold{P})) \to
K_0(C(\bold{P}^1(\bold{R}) \times 
\bold{P})\rtimes \bold{Z}/2 ) \oplus K_0(C(\bold{P}^1(\bold{R}) \times
\bold{P})\rtimes \bold{Z}/3 ). 
$$
The kernel of this map satisfies
$$ 
Ker(\alpha) \cong H_1(X_G, \text{cusps}, \bold{Z}). 
\eqno(2.10) 
$$
In particular, the 
modular symbols (2.9) are identified with elements in $Ker(\alpha)$:
$$
 \{ g(0), g(i\infty) \}_G \leftrightarrow \delta_{s} -
\delta_{\sigma(s)}, 
\eqno(2.11) 
$$
where $\delta_{s}$ is the projector in 
$C(\bold{P}^1(\bold{R}) \times \bold{P})$ given by the function equal
to one on the sheet $\bold{P}^1(\bold{R}) \times \{ s \}$ and zero
elsewhere. 
\endproclaim

\smallskip

Via the six terms exact sequence, the elements of $Ker(\alpha)$ can
be identified with (the image of) elements in $K_0(C(\bold{P}^1(\bold{R})
\times \bold{P}) \rtimes \Gamma)$. Thus, modular symbols, that is,
homology classes of certain geodesics in the bulk space correspond to
(differences of) 
projectors in the algebra of observables on the boundary space.   

\medskip

{\bf 2.2.5. Modular forms.} Finally, we discuss from this perspective
some results of [ManMar], [Mar], which give a correspondence between
certain classes of functions on the bulk space and on the boundary.

\smallskip

As the class of functions on the boundary, we consider functions
$$ 
\ell(f,\beta)=\sum_{k=1}^\infty f(q_k(\beta),q_{k-1}(\beta)).
\eqno(2.12) 
$$
Here $f$ is a complex valued function defined on pairs of coprime integers
$(q,q')$ with $q\geq q'\geq 1$ and with $f(q,q') =O(q^{-\epsilon})$ for
some $\epsilon >0$, and $q_k(\beta)$ are the successive denominators of the
continued fraction expansion of $\beta \in [0,1]$.
The summing over pairs of successive denominators is what replaces
modularity, when ``pushed to the boundary''.

\smallskip

We consider the case of $G=\Gamma_0(N)$, and the function
$$
f(q,q')=\frac{q+q'}{q^{1+t}}
\int_{\left\{ 0, \frac{q'}{q} \right\}_{\Gamma_0(N)}}
\omega, 
\eqno(2.13) 
$$
with $\omega$ such that the pullback $\pi_G^*(\omega)/dz$ is an
eigenform for all Hecke operators.  
Consider the corresponding $\ell(f,\beta)$ defined as in (2.12). We
have the following result.

\smallskip

\proclaim{\quad 2.2.6. Proposition} For almost all $\beta$, the series
(2.12) for the function (2.13) converges absolutely. Moreover, we have
$$ 
C(f,\beta):= \sum_{n=1}^\infty \frac{q_{n+1}(\beta)+
q_n(\beta)}{q_{n+1}(\beta)^{1+t}}\,  
\left\{ 0,\frac{q_n(\beta)}{q_{n+1}(\beta)}\right\}_{\Gamma_0(N)}
\eqno(2.14) 
$$
which defines, for almost all $\beta$ a homology class in
$H_1(X_G,\text{cusps},\bold{R})$ satisfying
$$ \ell(f,\beta)= \int_{C(f,\beta)} \omega 
\eqno(2.15) 
$$
with integral average 
$$ 
\int_{[0,1]} \ell(f,\beta) \, d\beta = \left( \frac{\zeta
(1+t)}{\zeta (2+t)} 
-\frac{L_{\omega}^{(N)}(2+t)}{\zeta^{(N)} (2+t)^2}
\right) \,\int_0^{i\infty}\Phi(z) dz, 
\eqno(2.16) 
$$
with $L_{\omega}^{(N)}$ the Mellin transform of $\Phi$ with omitted
Euler $N$--factor, and $\zeta(s)$ the Riemann zeta, with
corresponding $\zeta^{(N)}$.
\endproclaim
 
\medskip

Results of this type  can be regarded, on the one hand, as an explicit
correspondence between a certain class of fields on the bulk space
(Mellin transforms of modular forms of weight two), and the class of
fields (2.12) on the boundary. It also provides classes (2.14)
which correspond to certain configurations of geodesics in the bulk
space. These can be interpreted completely in terms of the
boundary. In fact the results of Proposition 2.2.4 can be rephrased
also in terms of cyclic cohomology (cf.~[ManMar], [Nis]), so that the
classes (2.14) in $H_1(X_G,\text{cusps},\bold{R})$ can be regarded as
elements in the cyclic cohomology of the algebra (2.2). Thus,
expressions such as the right hand side of (2.16), which express
arithmetic properties of the modular curve can be recast entirely in
terms of a suitable field theory on the boundary (2.2).

\bigskip

\centerline{\bf References}

\medskip

[AhGuMOO] O.~Aharony, S.~Gubser, J.~Maldacena, H.~Ooguri,
Y.~Oz. {\it Large N field theories, string theory
and gravity.} Phys. Rep. 323 (2000), no. 3-4, 183--386.

\smallskip

[AlGo] E.~\'Alvarez, C.~G\'omez. {\it Geometric
holography, the renormalization group and the $c$--theorem}.
 Nuclear Phys. B 541 (1999), no. 1-2, 441--460.

\smallskip

[ABMNV] L.~Alvarez-Gaum{\'e}, J.-B.~Bost, G.~Moore, Ph.~Nelson, and
C.~Vafa, {\it Bosonization on higher genus Riemann surfaces}. 
Comm. Math. Phys. 112(3) (1987), 503--552.

\smallskip

[AMV] L.~Alvarez-Gaum{\'e}, G.~Moore, and C.~Vafa, 
{\it Theta functions, modular invariance, and strings}.
Comm. Math. Phys. 106(1) (1986), 1--40.
\smallskip

[ABBHP] S.~{\AA}minneborg, I.~Bengtsson, D.~Brill, S.~Holst, and 
P.~Peld{\'a}n, {\it Black holes and wormholes in $2+1$ dimensions}.  
Classical and Quantum Gravity, 15(3) (1998), 627--644.

\smallskip

[Ar] S.~Ju.~Arakelov, {\it An intersection
theory for divisors on an arithmetic surface}. Izv. Akad. Nauk SSSR
Ser. Mat. 38:1179--1192, 1974. 

\smallskip

[BR] V.~Balasubramanian, S.~Ross, {\it Holographic particle
detection}. Phys. Rev. D (3) 61 (2000), no. 4, 12 pp.

\smallskip

[BTZ] M.~Ba\~{n}ados, C.~Teitelboim, J.~Zanelli, {\it Black hole in
three-dimensional spacetime}. Phys. Rev. Lett. 69 (1992), no. 13,
1849--1851. 

\smallskip

[Be] L.~Bers, {\it Simultaneous uniformization.}. Bull. AMS
66 (1960), 94--97.

\smallskip

[BKSW] D.~Birmingham, C.~Kennedy, S.~Sen, and A.~Wilkins,
{\it Geometrical finiteness, holography, and the
  Ba\~nados--Teitelboim--Zanelli black hole}.
 Phys. Rev. Lett. 82(21) (1999), 4164--4167.

\smallskip

[Bow] R.~Bowen. {\it Hausdorff dimension of quasicircles}.
Inst. Hautes \'Etudes Sci. Publ. Math. 50, (1979), 11--25.

\smallskip

[BowSer] R.~Bowen, C.~Series, {\it Markov maps associated with
Fuchsian groups}. Inst. Hautes \'{E}tudes
Sci. Publ. Math. 50 (1979), 153--170.

\smallskip

[Br1] J.~F.~Brock. {\it The Weil--Petersson metric and volumes
of 3--dimensional hyperbolic convex cores.}
e--Print math.GT/0109048

\smallskip

[Br2] J.~F.~Brock. {\it Weil--Petersson translation distance and
volumes of mapping tori}, preprint arXiv:math.GT/0109050.

\smallskip

[ChMay] C.~H.~Chang, D.~H.~Mayer, {\it Thermodynamic formalism and
Selberg's zeta function for modular groups}. Regul. Chaotic Dyn. 
5(3) (2000), 281--312.

\smallskip

[Fay] J.~Fay. {\it Kernel functions, analytic torsion, and moduli
spaces.} Memoirs of the AMS, vol. 96, N 464, AMS, Providence RA, 1992.

\smallskip

[FeSo] F.~Ferrari, J.~Sobczyk,
{\it Bosonic field propagators on algebraic curves}.
J. Math. Phys. 41(9) (2000), 6444--6462.

\smallskip

[GvP] L.~Gerritzen, M.~van~der Put, {\it Schottky groups and Mumford
curves}. Springer, 1980.

\smallskip

[Kh] A.~Kholodenko. {\it Boundary conformal field theories,
limit sets of Kleinian groups and holography.}
J. Geom. Phys. 35, N. 2-3 (2000), 193--238.

\smallskip

[Kr1] K.~Krasnov, {\it Holography and Riemann surfaces}. Adv. Theor. Math. 
Phys. 4(4) (2000). 

\smallskip

[Kr2] K.~Krasnov, {\it 3D gravity, point particles and Liouville theory.}
e--Print arXiv:hep--th/0008253

\smallskip

[Kr3] K.~Krasnov, {\it On holomorphic factorization in asymptotically
AdS 3D gravity.} e--Print arXiv:hep--th/0109198 

\smallskip

[Kr4] K.~Krasnov, {\it Analytic continuation for asymptotically
AdS 3D gravity}. e--Print arXiv:gr--qc/0111049.

\smallskip

[Lew] J.~Lewis, {\it Spaces of holomorphic functions equivalent to the
even Maass cusp forms}. Invent. Math. 127(2) (1997), 271--306.

\smallskip 

[LewZa1] J.~Lewis, D.~Zagier, {\it Period functions and
the Selberg zeta function for the modular group.}
In: The Mathematical Beauty of Physics, Adv. Series in
Math. Physics 24, World Scientific, Singapore, 1997, pp. 83--97.

\smallskip

[LewZa2] J.~Lewis, D.~Zagier, {\it Period functions for Maass wave
forms. I}. Ann. of Math. (2) 153(1) (2001), 191--258.

\smallskip 

[Mal] J.~Maldacena. {\it 
The large $N$ limit of superconformal field theories and
  supergravity}. Adv. Theor. Math. Phys. 2(2) (1998), 231--252.

\smallskip

[MMS] J.~Maldacena, J.~Michelson, A.~Strominger. {\it Anti-de Sitter
fragmentation}. J. High Energy Phys. 1999, no. 2, Paper 11, 23 pp.

\smallskip

[Man1] Yu.~I.~Manin. {\it The partition function
of the Polyakov string can be expressed in terms
of theta--functions}. Phys. Lett. B 172:2 (1986), 184--185. 

\smallskip

[Man2] Yu.~I.~Manin. {\it Three-dimensional hyperbolic geometry as
$\infty$-adic Arakelov geometry}. Invent. Math. 104(2) (1991),
223--243.
 
\smallskip

[ManD] Yu.~I.~Manin, V.~Drinfeld, {\it Periods of $p$-adic {S}chottky
groups}, J. reine angew. Math. 262/263 (1973), 239--247.

\smallskip

[ManMar] Yu.~I.~Manin, M.~Marcolli, {\it Continued fractions, modular
symbols, and non-commutative geometry}. e--Print 
  arXiv:math.NT/0102006.

\smallskip

[Mar] M.~Marcolli, {\it Limiting modular symbols and the Lyapunov
spectrum}. e--Print arXiv:math.NT/0111093.

\smallskip

[May] D.~H.~Mayer, {\it Continued fractions and related
transformations.} In: Ergodic Theory, Symbolic Dynamics
and Hyperbolic Spaces, Eds. T.~Bedford et al., Oxford
University Press, Oxford 1991, pp. 175--222.

\smallskip

[Mc] C.~T.~McMullen, {\it From dynamics on surfaces to rational points
on curves}.  Bull. Amer. Math. Soc. (N.S.) 37(2) (2000), 119--140. 

\smallskip

[Mum] D.~Mumford, {\it An analytic construction of degenerating curves
over complete local rings}. Compositio Math. 24 (1974), 129--174.

\smallskip

[Nis] V.~Nistor, {\it Group cohomology and cyclic cohomology of
crossed products}, Invent. Math. 99 (1990), 411--424.

\smallskip

[Pim] M.~Pimsner. {\it $KK$--groups of crossed products by groups
acting on trees}, Invent. Math. 86 (1986), 603--634.

\smallskip

[RS] D.~B.~Ray and I.~M.~Singer.
{\it Analytic torsion for complex manifolds.}
 Ann. of Math. (2), 98 (1973), 154--177.

\smallskip

[Rez] A.~Reznikov, {\it Three-manifolds class field theory}. 
Selecta Math. (N.S.) 3(3) (1997), 361--399. 

\smallskip

[Str] A.~Strominger. {\it $AdS\sb 2$ quantum gravity and string
theory}. J. High Energy Phys. 1999, no. 1, Paper 7, 19 pp. 

\smallskip

[Sul] D.~Sullivan. {\it On the ergodic theory at infinity of an
arbitrary discrete group of hyperbolic motions}. In {\it 
Riemann surfaces and related topics: Proceedings of the 1978
  Stony Brook Conference},
  pp. 465--496, Princeton Univ. Press 1981. 

\smallskip

[Suss] L.~Susskind. {\it The world as a hologram}. 
J. Math. Phys. 36 (1995), no. 11, 6377--6396. 

\smallskip

[TaZo] L.~Takhtajan, P.~Zograf. {\it On uniformization
of Riemann surfaces and the Weyl--Petersson metric
on Teichm\"uller and Schottky spaces.}
Math. USSR Sbornik, 60 (1988), 297--313.

\smallskip

['tH] G. 't Hooft. {\it Dimensional Reduction in Quantum Gravity}. 
e--Print arXiv:gr-qc/9310026.

\smallskip

[We] A.~Werner. {\it Arakelov intersection indices
of linear cycles and the geometry of buildings
and symmetric spaces.} e--Print arXiv:math.AG/0006128.

\smallskip
 
[Wi] E.~Witten, {\it Anti de Sitter space and
holography}. Adv. Theor. Math. Phys. 2(2) (1998), 253--291.

\smallskip

[WiY] E.~Witten, S.--T.~Yau. {\it Connectedness of the
boundary in the AdS/CFT correspondence.} 
Adv. Theor. Math. Phys. 3(6) (1999), 1635--1655 .

\enddocument